# Differential Reprogramming Based on Constructive Interference for Wireless Sensor Network

Bing Hu

[1] *Key Laboratory of Broadband Wireless Communication and Sensor Network Technology, Nanjing University of Posts and Telecommunications, Ministry of Education, Nanjing 210003, China*

**Abstract:** To improve the performance of reprogramming in wireless sensor network, we present a novel reprogramming structure and constructive interference-based dissemination protocol (CIDP) to transmit the patch through out the network fast and reliability. CIDP disseminates the patch, which is divided into several packets, to the network exploiting constructive interference. We evaluate our implementation of CIDP using simulation under different number of nodes. Our results show that CIDP disseminates the patch less than 4 milliseconds. In general, the probability of a node receives the complete patch as high as 99.99%.

## 1. Introduction

Wireless sensor network is a multi-hop ad hoc network, which is computing with a large number of ubiquitous, low-power tiny sensor nodes. In general, wireless sensor networks are deployed in unattended for a long time, e.g., environment and habitat monitoring or military surveillance, they have to adapt either to the surroundings or to any demand changes of the network applications. In order to obtain this adaptability, network reprogramming is a necessary behavior during which new code or parameters are put onto the nodes wireless.

Wireless reprogramming is an active research filed, which have two components, reprogramming scheme and dissemination protocol. The objective of reprogramming schemes is to achieve near zero external flash writes, minimum internal flash writes, and small size patch, which should disseminate to the whole network. Dissemination protocols are used for fast, reliable and low-cost transfer of patch updates to nodes within the deployed WSNs.

There are numerous reprogramming schemes [1-5] and data dissemination protocols [6-10] in the literature. Existing reprogramming schemes can be divided into four categories:1) systematic reprogramming, 2) modular reprogramming, 3) virtual machine based reprogramming, 4) differential reprogramming. The performance of systematic, modular and virtual machine based reprogramming is poor because those approaches need to transmit large size patch and demand high processing time and capabilities to be provided by the node's CPU, it is intolerable for the energy-limited nodes. Differential approaches [3-5] generate a patch by comparing the new code with the old version at the base-station. The salient feature of this approach is that the base-station only transmits the differential patch. QDiff [3]is a differential reprogramming approach, which generates the patch with smallest size by using clone detection to keep largest similarity between the old and new programs and eliminating the effect of variable moving in a new re-organization way, without writing the external flash memory.

Data dissemination is a fundamental building block in WSNs. Existing protocols can be divided into two categories: epidemic approaches [7-10] and constructive interference based protocols [6, 11, 12]. Epidemic approaches all shares a common feature that they employ a MAC protocol like CSMA/CA or TDMA for contention resolution, and typically their dissemination times are in the order of minutes for disseminating full image in practical networks. In order to eliminate the need for contention resolution, a new data dissemination fashion is proposed, which allows multiple senders exploiting constructive interference to transmit an identical packet simultaneously, and still guarantees receivers decode the packet correctly.

In this paper, we propose a novel reprogramming structure and constructive interference based dissemination protocol (CIDP) to transmit the patch through out the network. In the reprogramming structure, we use the method of QDiff to generate patch for the different between new and old version programs at the base-station, and then transmits the patch to all the nodes in the network exploiting CIDP.

The rest of this paper is organized as follows. Section 2 presents the related work, followed by the detail of reprogramming structure in section 3. Section 4 provides the simulation results, and the conclusion in section 5.

## 2. Related Work

Many differential reprogramming approaches have been proposed during the past decade. Zephyr [5] reduces the patch size by establishing a jump table, which all call and jump instructions switch to their destination based on the table. This method not only reduces the patch size, the node's energy consumption and processing requirements are both improved. However, for WSN applications, which have more loops, the scheme shows poor performance. $R^2$ [4] uses an efficient implementation of dynamic loading and linking modules to reduce the patch size. Unlike Zephyr, $R^2$ instead of jump table using meta-data, which is regarding changeable information, and transmits the difference of this meta-data to all nodes in the network. But, the implementation of this method needs to write the whole flash memory and a very large amount of code flash memory to maintain the meta-data. QDiff [3] is a differential reprogramming approach, which generates the patch with smallest size by using clone detection to keep largest similarity between the old and new programs and eliminating the effect of variable moving in a new re-organization way, without writing the external flash memory. Research studies have shown that the energy consumption for transmitting a single bit and executing 1000 instructions is equal [13]. Therefore, the updated patch should as small as possible.

A data dissemination protocol is a fundamental service required for the deployment and maintenance of practically re-program sensor nodes in the field. Constructive interference based protocol is an emerging trend to trickle the problem of reliability. In the seminal work on Glossy [12], a new flooding architecture exploits constructive interference, showed that constructive interference is practical in wireless sensor networks. It observed that there is a high probability result in constructive interference if the interval among these concurrent transmissions of the same packet is less than 0.5 microseconds. The implementation of Glossy is able to meet this requirement and a small packet can be flooded to all nodes with deterministic delays. However, the reliability of constructive interference decrease significantly as the number of concurrent transmitters increases, which was first studied by Wang et al. [11]. They proposed Spine Constructive Interference based Flooding (SCIF) to mitigate the scalability problem, but the correctness of SCIF assumes many conditions that are hard to achieve in practice. In contrast, Splash [6] for handling the scalability problem is fully practical solution based on collection tree protocols, which create parallel pipelines effectively by using constructive interference.

Different from these previous works that study the reprogramming schemes and dissemination protocols separately. We propose a novel reprogramming structure, by integrating the properties of QDiff and CIDP, to minimize the patch size, shorten the dissemination time delay and improve the reliability.

## 3 Reprogramming structure

In this section, we describe a new reprogramming structure that extends QDiff by exploiting constructive interference-based dissemination protocol (CIDP) to disperse the patch through whole network. There are several major steps related to the structure, as outlined in Fig. 1. Those are receive new and old files from system administer and code database respectively, calculate the patch using the method of QDiff and store (old, new, patch) tuple in the version control database for future usage, compress and fragment patch, transfer the encoded fragment throughout the network in turns by exploiting CIDP, sensor nodes receive patch and store it in memory, when they completely received the patch, boot loader reads, parses, and applies that patch on existing firmware without restart the node.

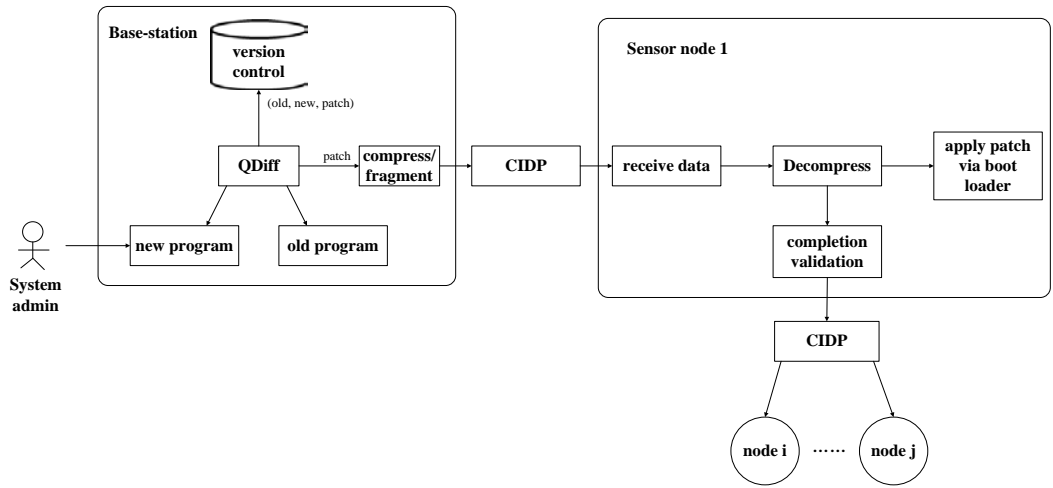

Fig. 1 A WSN reprogramming structure.

## 3.1 reprogramming scheme

Nasif proposed QDiff [3], a differential reprogramming approach, which utilizes clone detection to determine code changes efficiently. They handle branches, global variables, indirect addresses and relative branches by amending the ELF format in a manner, which is compatible with standard ELF. Fig. 2 shows the different between standard ELF file and amended ELF file. A key insight of QDiff is that it produces such a small patch and uploads the patch onto nodes without rewriting the flash memory and restarting the nodes.

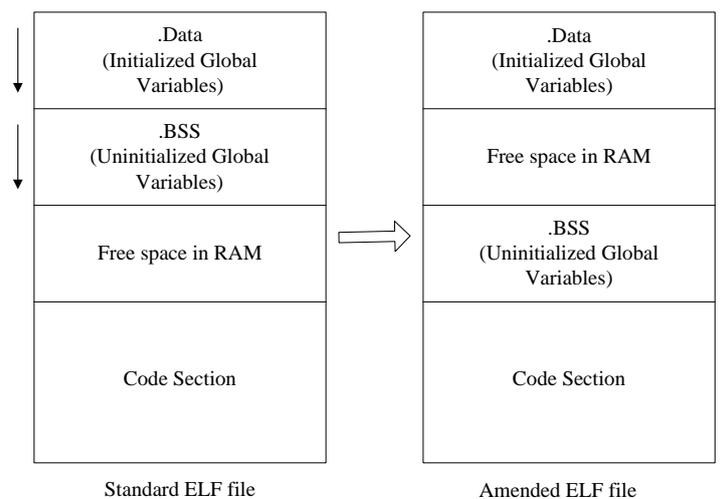

Fig. 2 Comparison of standard and amended ELF file.

QDiff considers executable files and the high-level source at the same time to generate similar part between old and new files employing clone detection tool, which calculates a list of function and variable clones between old and new files by taking in the C file of old and modified procedures. In order to reduce the differences between old and new procedures, QDiff separates the ELF files (both old and new)and reorders functions and global variables by installing new functions, uninitialized and initialized global variables at the end of code, bss and data sections respectively. In the standard ELF file, data and bss areas are both heap form. Because of this form, when add a new initialized variable in the data area all the variables in the bss area will be moved. This move will influence all the instructions associating with the modified variable and lead to a significantly large patch file. In the amended ELF file, bss section is organized in stack form, as shown in figure 2, and a free space is set between data and bss areas. Whether add or delete the initialized and uninitialized global variables, will not influent the variable addresses and the involving instructions. The base-station calculates the delta between the old and the new code with the above method and transmits it over the network using CIDP, which will be described in the next section.

**3.2 dissemination protocol**

We propose CIDP, a constructive interference based dissemination protocol, which is responsible for disseminating patch throughout the network fast and reliably. Because wireless is a broadcast medium, CIDP allows nodes overhear the wireless medium, and will forward overheard packets as soon as they receive them. Because the neighbors of a node receive a packet at the same time, they also forward the packet simultaneously. However, it is well known that the reliability of constructive interference decreases as packet size increases. To reduce the patch size, CIDP divides the data object of size $L_{obj}$ into packets of a fixed size $L_{pkt}$, where $L_{obj} = M \cdot L_{pkt}$, $M$ is an unfixed number of packets, which is changed with $L_{obj}$, as shown in Figure 3.

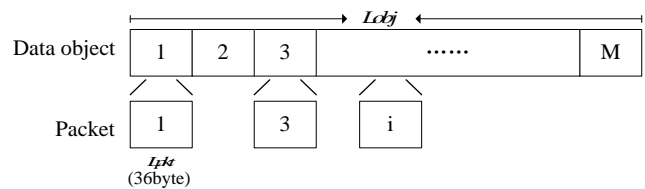

Fig. 3 Data fragment.

In order to know which packet will be received and whether the packets receive completely, we propose a format of CIDP packet with its default data payload size of 36 bytes (typically, communication packet is 36 bytes), i.e. $L_{pkt}$ is 36 bytes, which is depicted in Fig. 4.We record sequence number n, packets amount M and version number in the packet header, which represent the serial number of packet, the total number of packets that data object divide into and the version number for current update program respectively. Nodes can be repeatedly transmitting packet to improve dissemination reliability. We use $N_{max}$ to represent the maximum repeat times a node transmit a packet during a dissemination process, where a dissemination process is defined as a complete packet transmission through out the network with the same sequence number. After forwarding a packet, $n_{tx}$ is increased by 1 and compared to the maximum repeat times $N_{max}$. If $n_{tx}$ is equal to $N_{max}$, the dissemination is completes, then enter the next round, otherwise the node continue to relay packet.

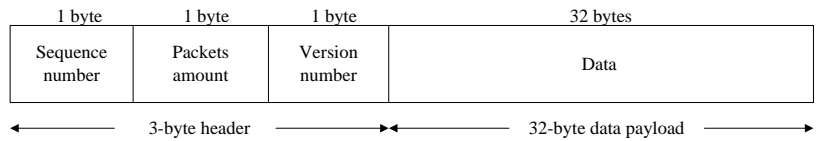

Fig. 4 Packet format used in CIDP.

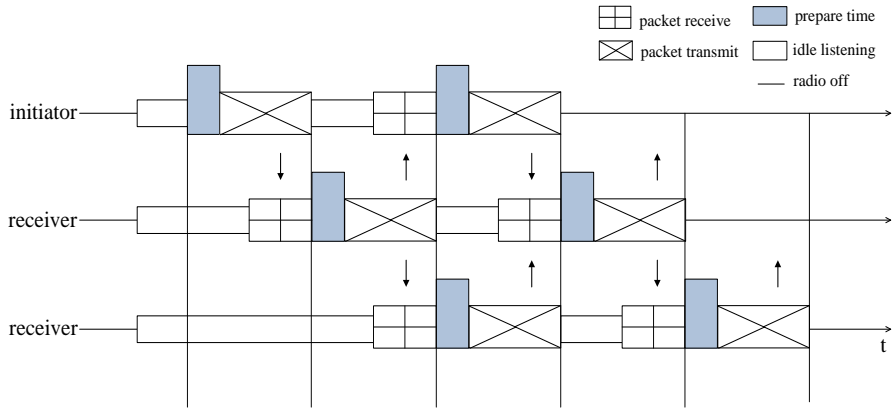

Fig. 5 Example of constructive interference-based dissemination with $N_{max} = 2$. Nodes always transmit the same packets.

Fig. 5 shows an example of a dissemination with $N_{max} = 2$. When CIDP starts, the initiator initializes $n_{tx}$ to 0 and triggers the first dissemination process by transmitting the first packet. After transmitting the packet, the initiator increase $n_{tx}$ to 1.Neighbors of the initiator overhear the packet and forward the packet immediately, and also increase $n_{tx}$ to 2. When their neighbors (including the initiator) overhear the packet, continue to forward the packet immediately, and so on. This process is end until $n_{tx}$ is equal to $N_{max}$ at all nodes, and then nodes start the second dissemination process until all the packets (i.e. sequence number is equal to packets amount) are transmitted completely.

## 4 Results of simulation

In this section, we analyze the simulation results of the proposed reprogramming structure with the methods of QDiff and CIDP. First, we present the patch size that QDiff generated, and then talk about the implementation of CIDF and the value of the maximum number of transmissions $N_{max}$. The simulation results for the structure this paper proposed will be discussed at last.

**QDiff.** The lower the patch size, the better the reprogramming scheme is. QDiff shown that it is outperformed Stream and Hermes by a factor of 250 and 12 times respectively. We can see from the experiment results, the largest patch size is less than 128 bytes, so, according to the method of 3.2, the patch can be divide into 4 packets at most. Besides, QDiff scheme outperformed both Stream and Hermes up to 21times in terms of internal flash usage, and don't required external flash memory to storing the golden image.

**CIDF.** The key requirement for constructive interference is that nodes have to transmit the same packet at the same time. We use the method that Glossy adopted to ensure that the nodes forwarding the received packet at the same time.

According to the results that the dissemination reliability increases almost logarithmically with $N_{max}$, and consistently exceeds 99% in a 5 hops network for N=92, and the dissemination latency is largely independent of $N_{max}$, shown in [12]. We set $N_{max} = \lfloor log_2(N + 1) \rfloor$, $N$ is the total number of nodes in the network. For the instance of 5 hops network, according to the experimental results of Glossy, the dissemination reliability for an 8-byte packet reached nearly 99.999%.

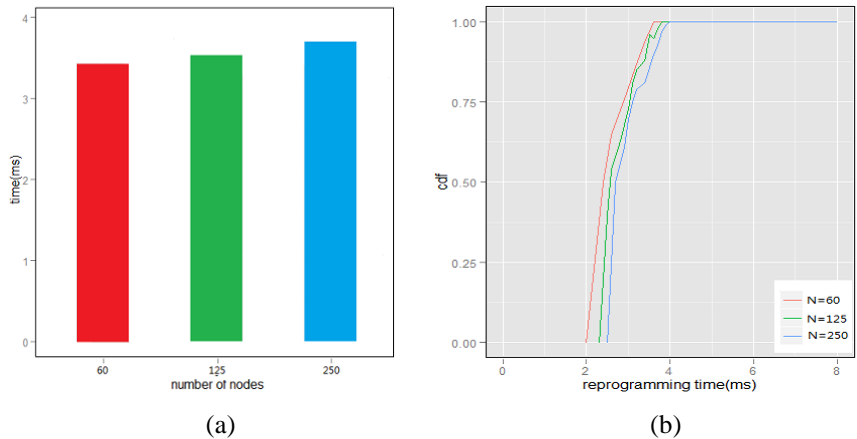

(a)      (b)

Fig.6 Reprogramming time for various N.

Next, we analyze the results from simulation based on network dissemination with a 128-byte patch. This patch is divided into 4 packets. Fig. 6(a) plots the average time for different $N$. We can see from Fig. 6(a) that CIDP disseminates a 128-byte patch to all nodes only needs less than 4ms. Fig. 6(b) presents the CDF of reprogramming time.

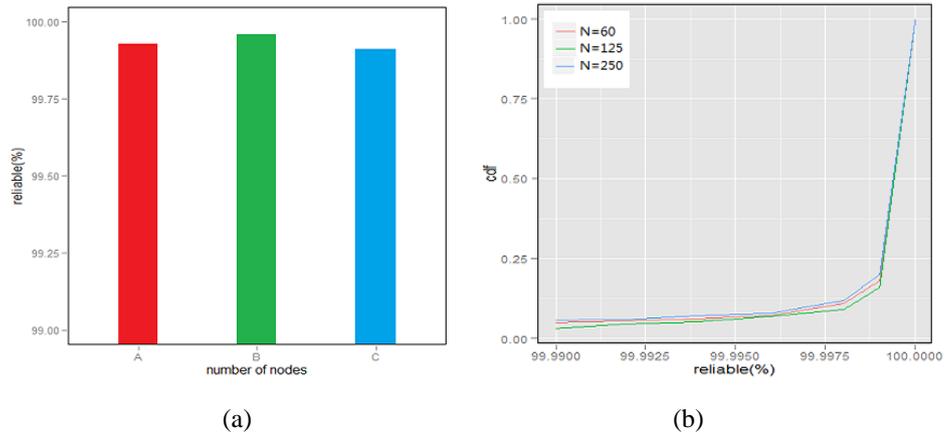

(a) (b)

Fig.7 Dissemination reliable for various N.

Looking at Fig. 7(a), it shows the dissemination reliability with different $N$. It is easy to observe that all nodes have dissemination reliability as high as 99.99% at the maximum number of nodes.Fig. 7(b) plots the CDF of dissemination reliability.

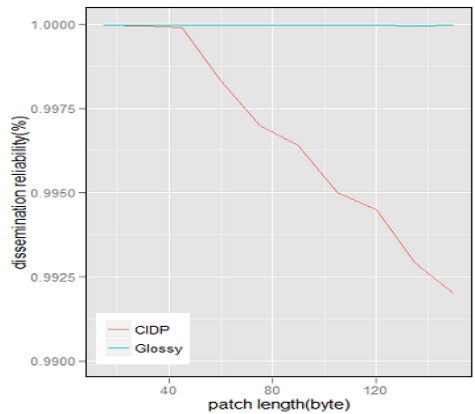

Fig.8 Dissemination reliability versus patch length L (N=94).

From Fig.8, we find that CIDP is superior to Glossy in terms of the dissemination reliability. Fig.8 shows dissemination reliability versus packet length L with fixed number of nodes N=94, so $N_{max} = 6$. As L increases, the dissemination reliability of Glossy decreases logarithmically while that of CIDP almost keeps the same. Particularly, when L=128, the dissemination reliability of Glossy is 99.34% while the dissemination reliability of CIDP is above 99.99%.

## 5 Conclusions

The purpose of this paper is to realize a fast and reliable reprogramming method. It is easy to find that such method would greatly benefit from efficient reprogramming scheme and fast dissemination protocol. This paper thus proposes a reprogramming structure and constructive interference-based dissemination protocol (CIDP) for wireless sensor networks. CIDP improve reprogramming reliability through patch fragmentation. We have evaluated our implementation of CIDP using simulation under different number of nodes. The results demonstrate that CIDP can provide fast and reliable dissemination for the patch.